\documentclass[aps,prd,superscriptaddress,12pt,showpacs,notitlepage]{revtex4}
\usepackage{amsmath,amssymb,mathrsfs}
\usepackage{graphicx}
\usepackage{subfig}

\newcommand{\MeV}{\textrm{MeV}}

\newcommand{\be}{\begin{equation}}
\newcommand{\ee}{\end{equation}}
\newcommand{\bea}{\begin{eqnarray}}
\newcommand{\eea}{\end{eqnarray}}
\newcommand{\pa}{\partial}

\begin{document}
\title{BPS Skyrmions as neutron stars}
\author{C. Adam}
\affiliation{Departamento de F\'isica de Part\'iculas, Universidad de Santiago de Compostela and Instituto Galego de F\'isica de Altas Enerxias (IGFAE) E-15782 Santiago de Compostela, Spain}
\author{C. Naya}
\affiliation{Departamento de F\'isica de Part\'iculas, Universidad de Santiago de Compostela and Instituto Galego de F\'isica de Altas Enerxias (IGFAE) E-15782 Santiago de Compostela, Spain}
\author{J. Sanchez-Guillen}
\affiliation{Departamento de F\'isica de Part\'iculas, Universidad de Santiago de Compostela and Instituto Galego de F\'isica de Altas Enerxias (IGFAE) E-15782 Santiago de Compostela, Spain}
\author{R. Vazquez}
\affiliation{Departamento de F\'isica de Part\'iculas, Universidad de Santiago de Compostela and Instituto Galego de F\'isica de Altas Enerxias (IGFAE) E-15782 Santiago de Compostela, Spain}
\author{A. Wereszczynski}
\affiliation{Institute of Physics,  Jagiellonian University,
Reymonta 4, Krak\'{o}w, Poland}


\begin{abstract}
The BPS Skyrme model has been demonstrated already to provide a physically intriguing and quantitatively reliable description of nuclear matter. Indeed, the model has both the symmetries and the energy-momentum tensor of a perfect fluid, and thus represents a field theoretic realization of the "liquid droplet" model of nuclear matter. In addition, the classical soliton solutions together with some obvious corrections (spin-isospin quantization, Coulomb energy, proton-neutron mass difference) provide an accurate modeling of nuclear binding energies for heavier nuclei.  
These results lead to the rather natural proposal to try to describe also neutron stars by the BPS Skyrme model coupled to gravity. We find that the resulting self-gravitating BPS Skyrmions provide excellent results as well as some new perspectives for the description of bulk properties of neutron stars when the parameter values of the model are extracted from nuclear physics. Specifically, the maximum possible mass of a neutron star before black-hole formation sets in is a few solar masses, the precise value depending on the precise values of the model parameters, and the resulting neutron star radius is of the order of 10 km.

\end{abstract}

\maketitle

\section{Introduction} 
The calculation of physical observables of strongly interacting matter at low energies - relevant, e.g., for nuclear physics - directly from QCD is a notoriously difficult problem, which led to the introduction of low-energy effective field theories (EFTs) as a more tractable alternative. The Skyrme model is a well-known example of such a low-energy EFT. It was introduced originally by Skyrme \cite{skyrme} as a purely mesonic nonlinear field theory
for the description of nuclei. Skyrme's idea was that nucleons should be described as a kind of "vorticity" in a mesonic "fluid" or, in a more modern language, as topological solitons of the underlying mesonic nonlinear field theory. And, indeed, the Skyrme model is known to possess topological solitons ("Skyrmions") whose topological index is identified with the baryon number. The original idea of Skyrme gained further support when it was observed that QCD in the limit of a large number of colors (large $N_c$) becomes a theory of weakly interacting mesons (interaction strength $\sim N_c^{-1}$) \cite{thooft}. Such weakly interacting nonlinear field theories frequently possess solitonic solutions with soliton masses proportional to the inverse of the (weak) coupling, which in the present case of the large $N_c$ mesonic model of QCD are identified with baryons and nuclei, recovering thereby the proposal of Skyrme.

The Skyrme model has been applied to the description of nuclei with notable success, e.g., in the description of rotational excitation bands of some light nuclei \cite{wood}, \cite{lau}. The version of the model originally proposed by Skyrme, however, has some drawbacks in the description of physical nuclei.  First of all, Skyrmions with higher baryon number $B$ have rather high binding energies (i.e., masses significantly below $B$ times the $B=1$ Skyrmion mass, see, e.g., \cite{manton-book}), which is in striking contrast to the low binding energies of physical nuclei. Also, Skyrmions for large baryon number tend to form crystals of lower $B$ substructures \cite{crystal1}, \cite{crystal2}, which is at odds with the liquid-type behaviour of physical heavy nuclei. These problems recently led to propose several "near BPS" Skyrme models, that is, generalizations of the original Skyrme model which are close to BPS models \cite{BPS-Sk}, \cite{sutcliffe}. Here by a BPS model we understand a field theory which has both an energy bound for static field configurations which is exactly linear in the baryon charge $B$ and solutions saturating the bound for all values of $B$ (we shall assume $B\ge 0$ in the sequel, i.e., consider only matter not antimatter). The original Skyrme model is not BPS. It has a lower topological energy bound, but it may be shown easily that this bound cannot be saturated. Specifically, we consider the following near BPS Skyrme model \cite{BPS-Sk} (for the moment in flat Minkowski space; we use the "mostly minus" metric sign convention ${\rm diag} (g_{\mu\nu}) = (+,-,-,-)$),
\be \label{near-BPS}
{\cal L}={\cal L}_0 + {\cal L}_6 + \epsilon ( {\cal L}_2 + {\cal L}_4) ,
\ee
where 
\be
{\cal L}_2 = - \lambda_2 {\rm tr}\; L_\mu L^\mu \; ,
\quad {\cal L}_4 = \lambda_4 {\rm tr}\; \left( [L_\mu ,L_\nu ] \right)^2
\ee
and 
\be
{\cal L}_0 = -\lambda_0 {\cal U}({\rm tr}\; U) \; , \quad {\cal L}_6 = -\lambda_6 \left( \epsilon^{\mu\nu\rho\sigma} {\rm tr}\; L_\nu L_\rho L_\sigma \right)^2 
\equiv -(24 \pi^2)^2 \lambda_6 {\cal B}_\mu {\cal B}^\mu .
\ee
Here $U: {\mathbb R}^3 \times {\mathbb R} \to {\rm SU(2)}$ is the Skyrme field, $L_\mu = U^\dagger \pa_\mu U$ is the left-invariant Maurer-Cartan current and ${\cal U}$ is a potential. The $\lambda_n$ are dimensionful, non-negative coupling constants, and ${\cal B}_\mu$ is the topological or baryon number current giving rise to the topological degree (baryon number) $B\in {\mathbb Z}$,
\be \label{top-curr}
{\cal B}^\mu = \frac{1}{24 \pi^2 } \epsilon^{\mu\nu\rho\sigma} {\rm tr}\; L_\nu L_\rho L_\sigma \; ,\quad B= \int d^3 x {\cal B}^0 .
\ee
${\cal L}_2 + {\cal L}_4$ is the model originally considered by Skyrme, and the above generalization is essentially the most general model which is both Poincar\'e invariant and no more than quadratic in first time derivatives, such that a standard hamiltonian can be found. This generalized Skyrme model is near BPS for sufficiently small values of the dimensionless parameter $\epsilon$, because the submodel
\be
{\cal L}_{06} = {\cal  L}_0 + {\cal L}_6 
\ee
is BPS. That is to say, the static energy functional $E_{06}[U]$ has an energy bound linear in $B$ and (in fact, infinitely many) minimizing field configurations saturating the bound for each $B$, \cite{BPS-Sk}. Further, this energy functional is invariant under volume-preserving diffeomorphisms (VPDs) on physical space, which are the symmetries of a perfect fluid. The energy-momentum tensor of the model ${\cal L}_{06}$ is, in fact, the energy-momentum tensor of a perfect fluid, as we shall see below.  These findings lead to the intriguing hypotesis that the near-BPS Skyrme model (\ref{near-BPS}) might be the correct low-energy EFT for the description of nuclear matter, as the BPS submodel ${\cal L}_{06}$ already provides a rather good description of some of its static properties. Indeed, the BPS Skyrme model allows for a very accurate description of nuclear binding energies \cite{bind}, \cite{marleau}, especially for heavy nuclei. It is the purpose of the present letter to couple the BPS Skyrme model to gravity and to use the resulting self-gravitating BPS Skyrmions for the description of neutron stars. 

We remark that there already exist several attempts to describe neutron stars using the original Skyrme model. In \cite{bizon} the hedgehog ansatz for higher $B$ was coupled to gravity but it turned out that - as in the non-gravitating case - higher $B$ hedgehogs are not stable. In \cite{piette1}, \cite{piette2} approximate Skyrmion configurations based on rational maps were used. Probably the most promising attempt within this context is using Skyrmion crystals \cite{walhout}, \cite{piette3}  because Skyrmion crystals are the true minimizers of the original Skyrme model for large $B$. The crystal structure is, however, at odds with the fact that, most likely, the core of neutron stars is in a superfluid phase. Also, full numerical calculations are not possible in this case such that certain assumptions about the right equation of state of Skyrme crystals under strong gravitational fields must be made. An accessible review can be found in \cite{nelmes-thesis}.

\section{BPS Skyrme model and parameter values}
Conveniently redefining the coupling constants $\lambda_6 = \lambda^2 /(24)^2$ and $\lambda_0 = \mu^2$,
the static energy functional of the theory is
\be \label{BPS-en-funct}
E_{06} = \int d^3 x \left( \pi^4 \lambda^2  {\cal B}_0^2 + \mu^2 {\cal U}({\rm tr}\; U) \right) .
\ee
Its BPS bound
\be
E_{06} \ge  2\pi^2 \lambda \mu |B| \langle\sqrt{{\cal U}}\rangle_{{\mathbb S}^3} \; , \quad \langle\sqrt{{\cal U}}\rangle_{{\mathbb S}^3} \equiv \frac{1}{2\pi^2} \int_{{\mathbb S}^3} d\Omega \sqrt{{\cal U}}
\ee
(where $\langle\sqrt{{\cal U}}\rangle_{{\mathbb S}^3}$ is the average value of $\sqrt{{\cal U}}$ on the target space SU(2) $\sim {\mathbb S}^3$) is saturated by infinitely many BPS solutions \cite{BPS-Sk}, \cite{fosco}, \cite{speight}, and the corresponding BPS equation is
\be \label{BPSeq}
\pi^2 \lambda {\cal B}_0 \pm \mu \sqrt{{\cal U}} =0.
\ee
We now have to determine the values of the parameters $\lambda$ and $\mu$ to be used in our calculations.
The product ${\bf m}\equiv \lambda \mu $ has the dimensions of mass (energy; we use units where the speed of light $c=1$). Further, ${\bf l}\equiv (\lambda /\mu )^{1/3}$ has the dimensions of length. We fit ${\bf m}$ by requiring that the BPS Skyrmion mass is $B$ times one-fourth of the mass of the helium nucleus, $E_{06} = B\bar m_{\rm N}$ where $\bar m_{\rm N} =m_{\rm He}/4 = 931.75\; \MeV$. We use $\bar m_{\rm N}$ instead of the nucleon mass $m_{\rm N} \sim 940\; \MeV$ because the latter will receive contributions from (iso)spin excitations in a Skyrme model description, but these are absent for helium. Even helium receives small (e.g., coulombic) contributions in addition to the Skyrmion mass, but the uncertainty will be at most a few \MeV. To fix ${\bf l}$, we use the fact that BPS Skyrmions for many potentials (in particular, for the potentials considered in this letter) are compactons with a strictly finite volume $V$, and this volume is the same for all solutions with a given baryon number $B$ and is exactly linear in $B$. This permits to define a Skyrmion radius $R$ via $V=(4\pi/3) R^3$. We now require that this radius coincides with the nucleon radius $r_{\rm N} = 1.25\; {\rm fm}$ for $B=1$, i.e., $R=r_{\rm N} B^{1/3}$. We think that the fit for the mass parameter ${\bf m}$ is quite accurate, because by far the biggest contribution to the nuclear masses must always come from the Skyrmion mass. On the other hand, the fit for the length parameter ${\bf l}$ is probably less precise. Firstly, although the compacton radius is quite natural, there are additional definitions for radii (diverse charge radii) which could be used. For compactons these charge radii are always smaller than the compacton radius, indicating that the latter could be slightly bigger than the nucleon radius. Secondly, going  beyond the BPS submodel by including, e.g., the Dirichlet term  ${\cal L}_2$, the effects of the pion cloud will tend to increase the radius, indicating that the compacton radius $R$ without pion cloud could be somewhat smaller. To summarize, although our simple fit for ${\bf l}$ certainly provides a reasonable value, the true best fit value could easily deviate about 20\% or 30\% in either direction. Determining this true value, however, requires the knowledge of the complete low-energy EFT with all terms (also the non-BPS ones) included, which is beyond the scope of this letter.

Concretely, we shall consider the pion mass potential ${\cal U}_\pi = 1-\cos \xi$ and the pion mass potential squared  ${\cal U}_4 = {\cal U}_\pi ^2 $ with a quartic behaviour near the vacuum (here $U=\exp (\xi \vec n\cdot \vec \tau )$, $\vec n^2 =1$ and $\vec \tau$ are the Pauli matrices) with energies, compacton radii and fit values 
\be \label{param-pot-2}
{\cal U}_\pi: \; \;  E_{06} = \frac{64 \sqrt{2}\pi}{15} B\lambda \mu  ,  \; \;  R = \sqrt{2}\left(\frac{\lambda B}{\mu}\right)^\frac{1}{3} \; \; 
 \Rightarrow \; \;  {\bf m} = 49,15 \; \MeV  , \quad {\bf l} = 0,884 \; {\rm fm}  
\ee
\be \label{param-pot-4}
{\cal U}_\pi^2: \; \;  E_{06} = 2\pi^2 B \lambda \mu ,  \; \;  R = \left( \frac{3\pi B}{2}\right)^\frac{1}{3} \left( \frac{\lambda}{\mu}\right)^\frac{1}{3} \;  \; 
  \Rightarrow \; \;  {\bf m} = 47.20 \; \MeV  , \; \;  {\bf l}= 0.746 \; {\rm fm} 
\ee
These expressions for the energies and compacton radii may be calculated directly from the potentials, see \cite{thermo} (knowledge of the Skyrmion solutions is not required). 

\section{BPS Skyrmions coupled to gravity} 
The action of the BPS Skyrme model in a general metric $g_{\rho \sigma}$ (here $g= {\rm det} g_{\rho\sigma}$),
\be
S_{06} = \int d^4 x |g|^\frac{1}{2} \left( -\lambda^2 \pi^4 |g|^{-1} g_{\rho\sigma} {\cal B}^\rho {\cal B}^\sigma - \mu^2 {\cal U} \right) ,
\ee
leads to the energy-momentum tensor (${\cal B}^\rho$ is defined in Eq. (\ref{top-curr}))
\be
T^{\rho\sigma} = -2|g|^{-\frac{1}{2}}\frac{\delta}{\delta g_{\rho\sigma}} S_{06} = 2\lambda^2 \pi^4 |g|^{-1} {\cal B}^\rho {\cal B}^\sigma - \left( \lambda^2 \pi^4 |g|^{-1} g_{\pi\omega} {\cal B}^\pi {\cal B}^\omega - \mu^2 {\cal U} \right) g^{\rho\sigma} ,
\ee
which is the energy-momentum tensor of a perfect fluid (the perfect-fluid property of the term ${\cal L}_6$ alone, as well as its coupling to gravity, have already been discussed in \cite{slobo}), 
\be
T^{\rho \sigma} = (p+\rho )u^\rho u^\sigma - pg^{\rho\sigma}
\ee
where the four-velocity $u^\rho$, energy density $\rho$ and pressure $p$ are
\be 
u^\rho = {\cal B}^\rho / \sqrt{g_{\sigma \pi} {\cal B}^\sigma {\cal B}^\pi}
\ee
\bea
\rho &=& \lambda^2 \pi^4 |g|^{-1} g_{\rho\sigma} {\cal B}^\rho {\cal B}^\sigma + \mu^2 {\cal U} \nonumber \\
p &=& \lambda^2 \pi^4 |g|^{-1} g_{\rho\sigma} {\cal B}^\rho {\cal B}^\sigma - \mu^2 {\cal U} .
\eea
In the static case, and for a diagonal metric (which is sufficient for our purposes) we have $u^\rho = (\sqrt{g^{00}},0,0,0)$ and 
\be
T^{00} =\rho g^{00} \; ,\quad T^{ij} = -pg^{ij}.
\ee
In the flat space case, e.g., this implies that the pressure must be constant (zero for BPS solutions, nonzero for non-BPS static solutions \cite{thermo}), as a consequence of energy-momentum conservation,
\be
D_\rho T^{\rho\sigma} \to \partial_i T^{ij} =\delta^{ij}\partial_i p = 0, 
\ee
whereas $\rho$ will be a nontrivial function of the space coordinates.
In general, $\rho$ and $p$ will be quite arbitrary functions of the space-time coordinates, so there does not exist a universal equation of state (EoS)$p=p(\rho)$ which would be valid for all solutions.  

We now want to couple the BPS Skyrme model to gravity and solve the resulting Einstein equations for a static, spherically symmetric metric which in standard Schwarzschild coordinates reads
\be
ds^2 = {\bf A}(r) dt^2 - {\bf B}(r) dr^2 - r^2 (d\theta^2 + \sin^2 \theta d\phi^2 ).
\ee
For us the following observation is crucial. The above ansatz for the metric together with the axially symmetric ansatz for the Skyrme field with baryon number $B$
\be \label{axi-sym}
 \xi = \xi (r), \quad \vec n = (\sin \theta \cos B\phi , \sin \theta \sin B \phi , \cos \theta )
\ee
leads to a baryon density ${\cal B}^0$, energy density $\rho$ and pressure $p$ which are functions of $r$ only.
The ansatz is, thus, compatible with the Einstein equations
\be
G_{\rho\sigma} = \frac{\kappa^2}{2} T_{\rho \sigma} 
\ee
(here $G_{\rho\sigma}$ is the Einstein tensor and $\kappa^2 = 16 \pi G = 6.654 \cdot 10^{-41} \; {\rm fm} \; \MeV^{-1}$) and the static Euler-Lagrange equations for the Skyrme field, and reduces these equations to a system of ordinary differential equations (ODEs) in the variable $r$ for the three unknown functions ${\bf A}(r)$, ${\bf B}(r)$ and $\xi (r)$.  Before presenting this system of ODEs and the results of a numerical integration, 
we want to make some comments. 

Firstly, in flat Minkowski space the same axially symmetric ansatz (\ref{axi-sym}) (but referring to spherical polar coordinates in that case) was used in the calculations of nuclear binding energies in \cite{bind}. As said, the resulting binding energies are very accurate for heavier nuclei, but, nevertheless, once additonal terms (like, e.g., the Dirichlet term $\epsilon E_2$) are taken into account, there are strong arguments indicating that the  axially symmetric BPS Skyrmions are not the adequate ones (they do not minimize $E_2$ among all BPS Skyrmions) \cite{speight2}. An improved calculation using the true minimizers of $E_2$ and taking the contribution of $E_2$ into account should lead to even better results for the binding energies. Here we just want to emphasize that in the case of self-gravitating BPS Skyrmions the axially symmetric ansatz leading to a spherically symmetric metric, energy density and pressure {\em is} the correct one, essentially because gravity straightens out all deviations from spherical symmetry. Secondly, in the subspace of spherically symmetric solutions we may define a kind of EoS $p=p(\rho)$, because both $\rho$ and $p$ are functions of $r$. We find numerically that a simple power law
\be \label{EoS}
p = a \rho^b
\ee
reproduces this  EoS with a high precision. Here, however, $a$ and $b$ are not universal constants. Instead, they depend on the baryon number $B$. In particular, for "small" baryon number (small compared, e.g., to the solar baryon number $B_\odot$), where the effect of gravity may be neglected, the constant $a$ vanishes, $\lim_{B\to 0}a=0$ (the pressure is zero like in the case without gravity). If we treated the gravitational coupling $\kappa$ as a parameter which may vary then, of course, it would also hold that $\lim_{\kappa \to 0} a=0$. (Here we define $B_\odot$ as $B_\odot = M_\odot/\bar m_{\rm N} = 1.116 \cdot 10^{60} \MeV / 931.75 \; \MeV = 1.198 \cdot 10^{57}$, so strictly speaking $B_\odot$ is not the number of baryons in the sun, but the number of baryons (neutrons) in a neutron star with the same non-gravitational mass as the sun.)      

\section{Numerical results}
We find it convenient to introduce the new target space variable $h=(1/2)(1-\cos \xi )=\sin^2 \frac{\xi}{2}$ in what follows, with $h\in [0,1]$ and ${\cal U}_\pi = 2h$. The system of ODEs resulting from the Einstein equations may be brought into the form of a system of two equations for $h$ and ${\bf B}$, plus a third equation which determines ${\bf A}$ in terms of $h$ and ${\bf B}$. Explicitly, these equations read ($' \equiv \partial_r$)
\bea \label{eq1}
 \frac{1}{r}\frac{{\bf B}'}{{\bf B}} &=& - \frac{1}{r^2}({\bf B}-1) + \frac{\kappa^2}{2} {\bf B}\rho  \\ \label{eq2}
r\left( {\bf B}p\right)' &=& \frac{1}{2}(1-{\bf B}){\bf B}(\rho + 3 p) + \frac{ \kappa^2}{4} r^2 {\bf B}^2 (\rho - p) p  \\ \label{eq3}
\frac{{\bf A}'}{{\bf A}} &=& \frac{1}{r}({\bf B}-1) + \frac{\kappa^2}{2} r{\bf B}p
\eea
where $\rho$ and $p$ for the axially symmetric ansatz ($h_r \equiv \partial_r h$)
read
\be
\rho = \frac{4 B^2 \lambda^2}{{\bf B}r^4} h(1-h)h_r^2 + \mu^2 {\cal U}(h) , \quad p= \rho -2\mu^2 {\cal U}(h).
\ee
We integrate the system (\ref{eq1}), (\ref{eq2}) numerically via a shooting from the center. That is to say, we impose the boundary conditions $h(r=0)=1$ (anti-vacuum value) and ${\bf B}(r=0)=1$ (the amount of matter enclosed at $r=0$ is zero $\Rightarrow$ flat space metric). We are left with one free parameter, $h_2$, in the expansion about $r=0$, $h(r) \sim 1-(1/2) h_2 r^2 + \ldots$ or, equivalently, with $\rho_0 \equiv \rho(r=0) = B^2 \lambda^2 h_2^3 +\mu^2 {\cal U}(1)$. We then integrate from $r=0$ up to a point $r=R$ (compacton radius) where $h(R)=0$ (the vacuum). It follows easily from Eq. (\ref{eq2}) that, for a non-singular metric function ${\bf B}$, $p$ at $r=R$ must obey  $p' (R) =0$ which leads to a condition on $h_r (R)$, concretely
\be \label{cond}
\frac{4 B^2 \lambda^2}{{\bf B}(R) R^4} h_r^2 (R) -\mu^2 {\cal U}_h (0) =0.
\ee
In the numerical integration, the free parameter value $\rho_0$ is varied until this condition is met. Formal solutions which do not obey this condition produce metric functions ${\bf B}$ which are singular at $r=R$. In particular, such a metric function cannot be joined smoothly to the Schwarzschild solution in empty space (for $r\ge R$) and is, therefore, physically unacceptable. 

We find the following behaviour in the numerical integrations. For sufficiently small baryon number $B$, there exists precisely one "initial value" $\rho_0$ which obeys (\ref{cond}), i.e., one unique neutron star solution. For larger values of $B$ in a certain interval $B\in [B_*, B_{\rm max}]$, there exist {\em two} values for $\rho_0$ leading to solutions fulfilling condition (\ref{cond}). Interestingly, this is exactly like in the case of the Tolman-Oppenheimer-Volkoff (TOV) calculation where the neutrons are described by a free relativistic fermi gas (see e.g. \cite{weinberg}, Chapter 11.4, page 321).   
As in the TOV case, we assume that the lower value $\rho_0$ corresponds to the stable solution. Finally, for $B> B_{\rm max}$ solutions obeying condition (\ref{cond}) no longer exist. In other words, physically acceptable static solutions (neutron stars) which $B> B_{\rm max}$ do not exist. Instead, field configurations with such a large $B$ are unstable, indicating the collapse to a black hole. 

The neutron star solution found in the interval $r\in [0,R]$ is then smoothly joined to the vacuum solution for $r\ge R$. That is to say, $h(r)=0$ for $r\ge R$, and $B(r) = (1-\frac{2GM}{r})^{-1}$, from which the physical mass $M$ of the neutron star (with the gravitational mass loss taken into account) may be read off. 

One of the most important results is, of course, the value $B_{\rm max}$ and the corresponding maximal neutron star masses $M$ and radii $R$ for the two potentials ${\cal U}_\pi$ and ${\cal U}_\pi^2$ we consider. It is convenient to measure $B$ in solar units $n\equiv (B/B_\odot )$ (equivalently, $n = 
(B \bar m_{\rm N}/M_\odot)$, i.e., the non-gravitational mass of the baryon number $B$ skyrmion in solar mass units). Then, using the fit values (\ref{param-pot-2}) and (\ref{param-pot-4}), respectively,  we find for the maximum values
\be
{\cal U}_\pi: \quad n_{\rm max} = 5.005 , \quad M_{\rm max}  = 3.734 M_\odot , \quad R_{\rm max} = 18.458 \; {\rm km},
\ee
\be
{\cal U}_\pi^2 : \quad n_{\rm max} = 3.271 ,\quad M_{\rm max} = 2.4388 M_\odot , \quad R_{\rm max} = 16.801 \; {\rm km}.
\ee
We remark that neutron star masses up to about $M\sim 2M_\odot$ are firmly established, whereas there are indications for masses up to about $2.5 M_\odot$, see e.g. \cite{latt}, \cite{ozel} for an overview of recent measurements. The results of our calculations are, therefore, in excellent agreement with these observations, indicating that our model provides a very good description of the bulk properties of nuclear matter also in the presence of the gravitational interaction. Concerning the radii, we remark that the observational results are less precise. Besides, $R$ is the geometric radius which leads to a neutron star surface area of $4\pi R^2$, whereas when comparing to measurements sometimes other radii are more appropriate, like the proper distance from the origin to the surface, $\bar R = \int_0^R dr \sqrt{B(r)}$, or the radiation radius $R^* = R\sqrt{B(R)}$. Both $\bar R$ and $R^*$ are somewhat bigger than $R$ because ${\rm B}(r) \ge 1$. In any case, also our values for the radii are in the expected range of about $R\sim$ 10-20 km.
As said already, our fit for the unit of mass ${\bf m}$ is quite precise (determined by the nuclear mass $\bar m_{\rm N}$), but the unit of length ${\bf l}$ is less so, therefore it is interesting to study the  
sensitivity of both $M_{\rm max}$ and $R_{\rm max}$ under a change of the length scale,  ${\bf l} \to {\bf l'} = \alpha {\bf l}$. We find numerically that both $M_{\rm max}$ and $R_{\rm max}$ approximately change by a factor of $\alpha^{(3/2)}$ under this rescaling. 

Finally, we show our main numerical results in Figs. 1 - 4. Concretely, in Fig. 1a we plot the neutron star mass as a function of the non-gravitational Skyrmion mass, both in solar units. We find that for the extremal case $M_{\rm max}$ the gravitational mass loss is about 25\%. In Fig. 1b we plot $M$ against the (geometric) neutron star radius $R$. We find that even in the extremal case the neutron star radius is about a factor of two above the Schwarzschild radius. In Fig. 2 we show the equation of state for different values of $B$ (concretely for $n=1$ in Fig. 2a, and for $n_{\rm max}$ in Fig. 2b) together with the fit function $p=a\rho^b$ for appropriate values of $a,b$. 

In Fig. 3, we plot the metric function ${\bf B}(r)$ for several values of the baryon number $n=B/B_\odot$ close to its maximum value $n_{\rm max}$. We find for both potentials that the maximum value which ${\bf B}$ takes for $n=n_{\rm max}$ is about ${\bf B}_{\rm max} \sim 2.7$. It is interesting to compare this finding with the analogous result for the Skyrmion crystal of Ref. \cite{piette3}. There the authors calculated the minimum value of ${\bf B}^{-1}$ (which was called $S$ in that paper) for different solutions and always found that $S_{\rm min}>0.4$, which translates into ${\bf B}_{\rm max} <2.5$. So the ${\bf B}_{\rm max}$ we find for the maximum mass case is slightly bigger (i.e., the induced self-gravitation slightly stronger), but 
still quite similar to the result of \cite{piette3}. The position of the maximum of ${\bf B}(r)$ is quite close to the neutron star surface for the potential ${\cal U}_\pi$, whereas it is shifted towards the center for ${\cal U}_\pi^2$. This is related to the fact that, for ${\cal U}_\pi^2$, the energy density is more concentrated about the center (see Fig. 4).

In Fig. 4, we plot the energy densities for several values of the baryon number close to $n_{\rm max}$. We find that, especially for the potential ${\cal U}_\pi^2$, the energy density is quite sharply concentrated about the center. This may look surprising at first sight, but is simply related to the shape of the potential  ${\cal U}_\pi^2$, which is quite peaked about the anti-vacuum ($h=1$). Indeed, the BPS equation (\ref{BPSeq}) just states that the baryon density is proportional to the square root of the potential, so peaked potentials lead to peaked baryon density (and energy density) profiles already in the case without gravity. It is perhaps more instructive to compare the central energy density of the case without gravity to the central energy density for $n_{\rm max}$. The central energy density for the case without gravity does not depend on the baryon number $B$ and is given by $\rho_{\rm BPS}(r=0) = 2\mu^2 {\cal U}(h=1)$. Using the parameter values (\ref{param-pot-2}), 
we find for ${\cal U}_\pi$:  ${\rho_{\rm BPS}(r=0)} = 4({\bf m}/{\bf l}^3) = 285 \,{\rm MeV}\, {\rm fm}^{-3}$. The central energy density for $n_{\rm max}$ is, therefore, about 2.7 times the non-gravitational energy density $\rho_{\rm BPS}(r=0)$, see Fig. 4a. Similarly, we get for ${\cal U}_\pi^2$: 
$\rho_{\rm BPS}(r=0) = 8({\bf m}/{\bf l}^3) = 909 \, {\rm MeV}\, {\rm fm}^{-3}$.   In this case, the central energy density for $n_{\rm max}$ is just about 2.2 times the non-gravitational energy density $\rho_{\rm BPS}(r=0)$, see Fig. 4b. These results in both cases indicate a rather high stiffness of the effective (on-shell) EoS of strongly self-gravitating BPS Skyrmions, i.e., a nuclear matter which is only weakly compressible in strong gravitational fields. This result, again, compares quite well with the Skyrmion crystal results of Ref. \cite{piette3}, where a compression of the central energy density by not more than a factor of three is observed for all solutions. 

\begin{figure}[h]
 \begin{center} \hspace*{-1.0cm}
  \subfloat[$M/M_\odot$ vs. $B/B_\odot$]{\includegraphics[width=0.50\textwidth]{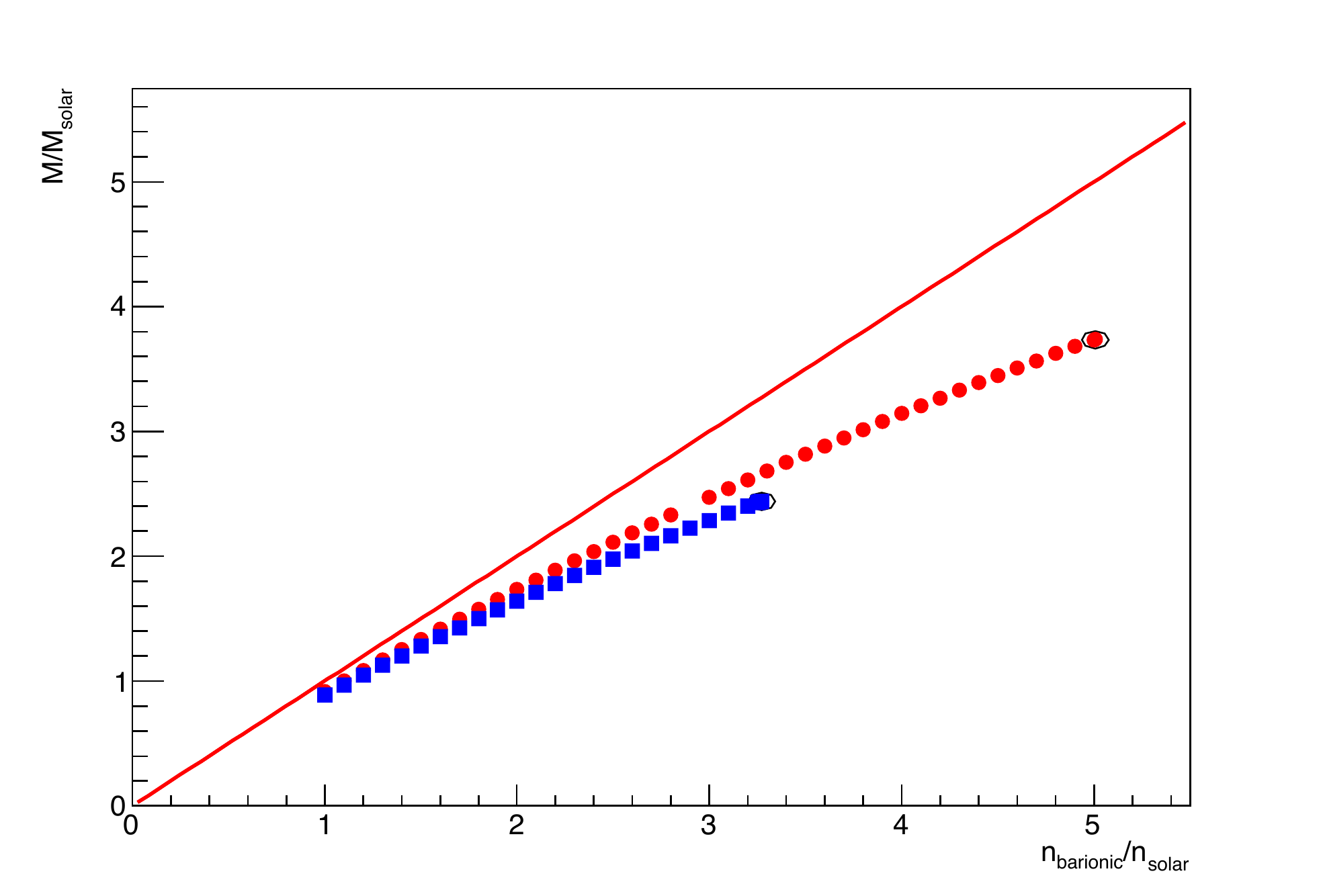}} \quad
  \subfloat[$M/M_\odot$ vs. $R$ (in km)]{\includegraphics[width=0.50\textwidth]{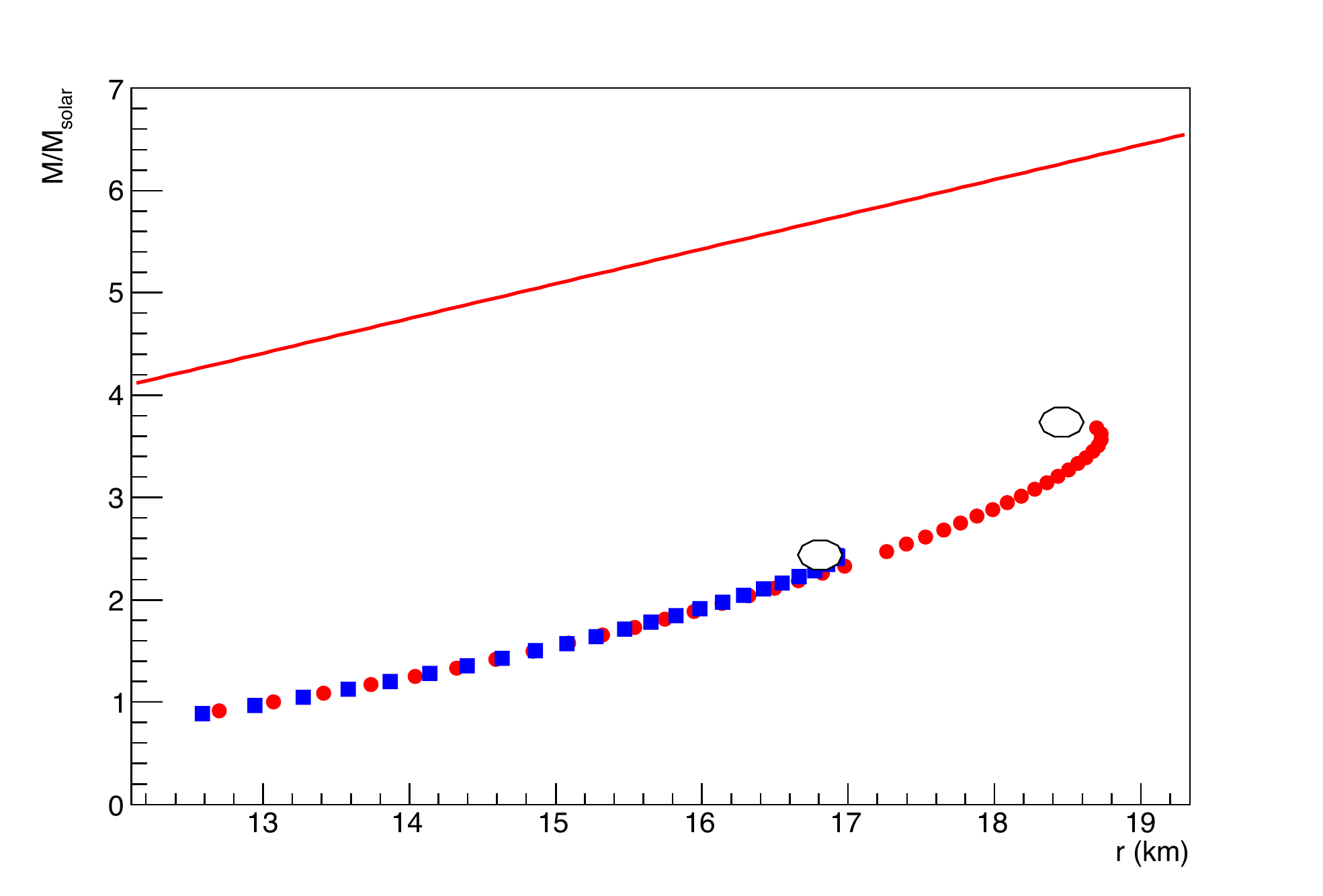}} \\
  \caption{1a) Neutron star mass as a function of baryon number, both in solar units. 
Symbol dot (red online): potential ${\cal U}_\pi$. Symbol square (blue online): potential ${\cal U}_\pi^2$. 
1b) Neutron star mass as a function of the neutron star radius. Potential ${\cal U}_\pi$:
symbol dot (red online). Potential ${\cal U}_\pi^2$: symbol square (blue online).
Maximum mass values are indicated by circles. The straight line is the Schwarzschild mass.}
  \label{fig1}
 \end{center}
\end{figure} 

\begin{figure}[h]
 \begin{center} \hspace*{-1.5cm}
  \subfloat[Equation of state for $n=1$]{\includegraphics[width=0.53\textwidth]{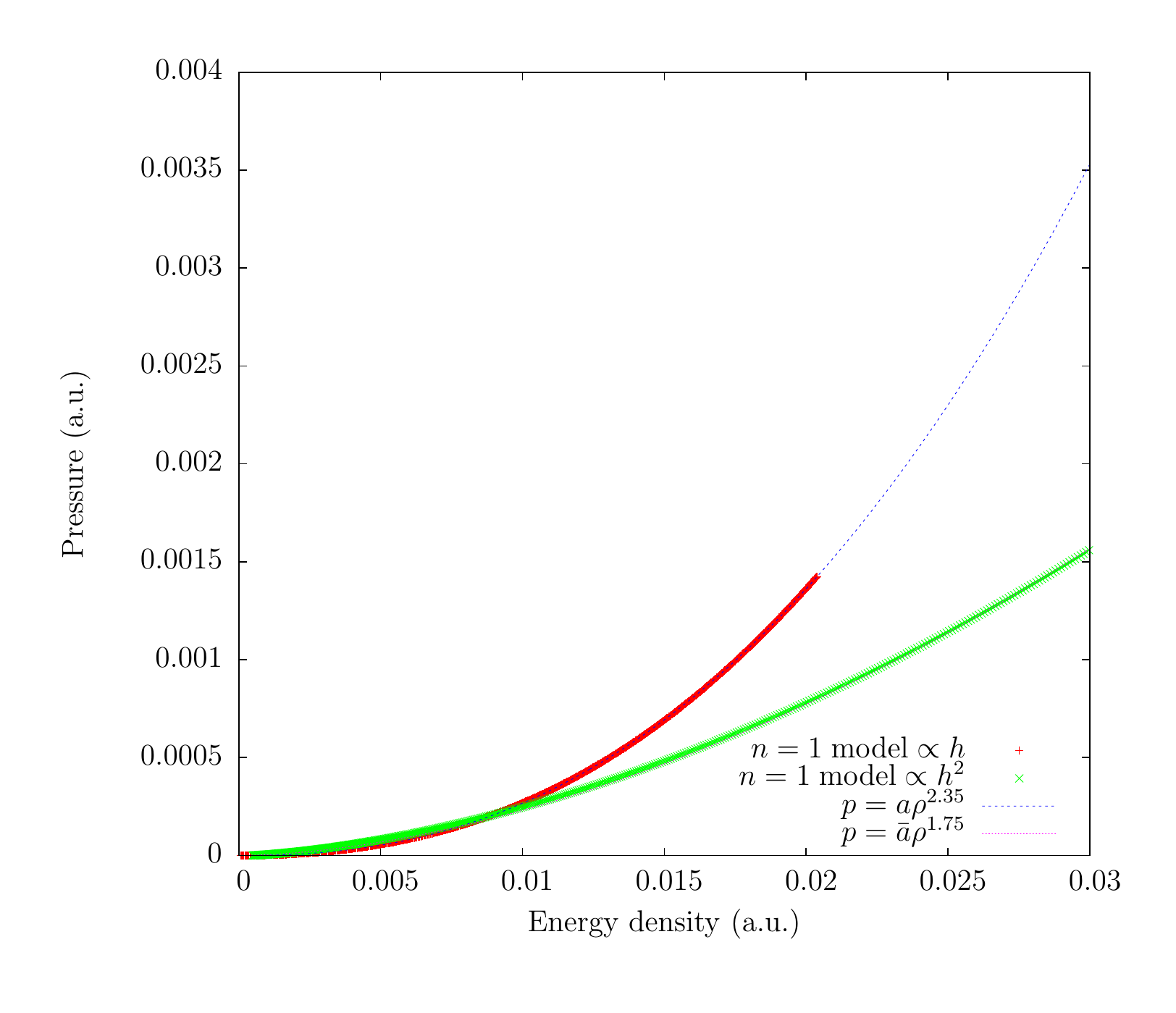}} \quad
  \subfloat[Equation of state for $n_{\rm max}$]{\includegraphics[width=0.53\textwidth]{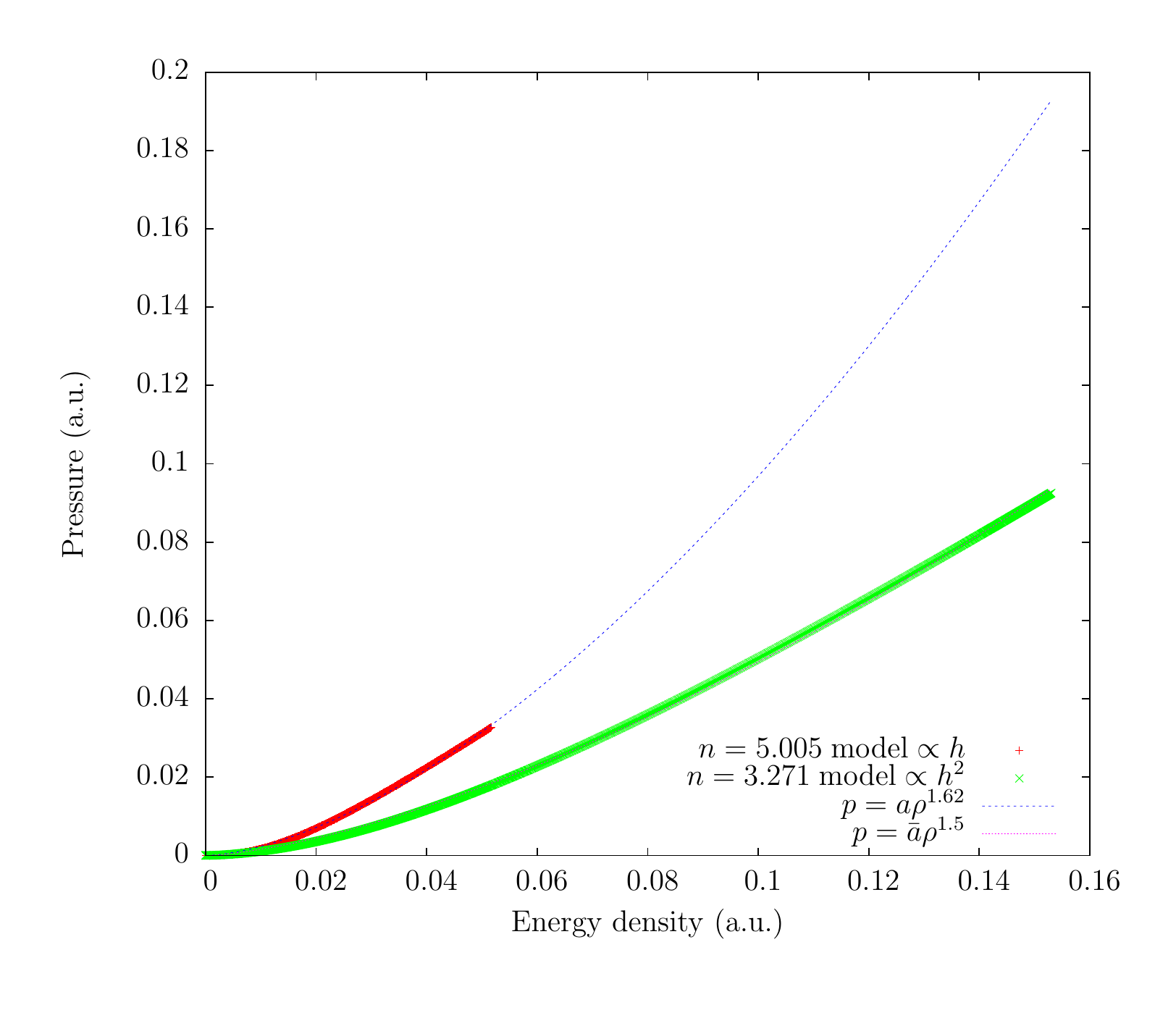}} \\
  \caption{Symbol plus ($+$, red online): potential ${\cal U}_\pi$. Symbol cross ($\times$, green online): potential ${\cal U}_\pi^2$. Dotted lines: corresponding fit functions.}
  \label{Old}
 \end{center}
\end{figure} 

\begin{figure}[h]
 \begin{center} \hspace*{-1.5cm}
  \subfloat[Metric function for ${\cal U}_\pi$]{\includegraphics[width=0.53\textwidth]{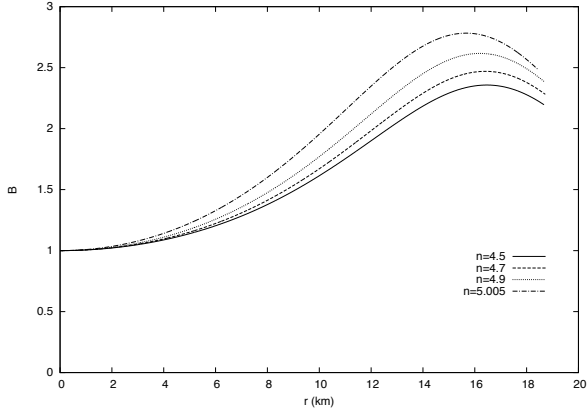}} \quad
  \subfloat[Metric function for ${\cal U}_\pi^2$]{\includegraphics[width=0.53\textwidth]{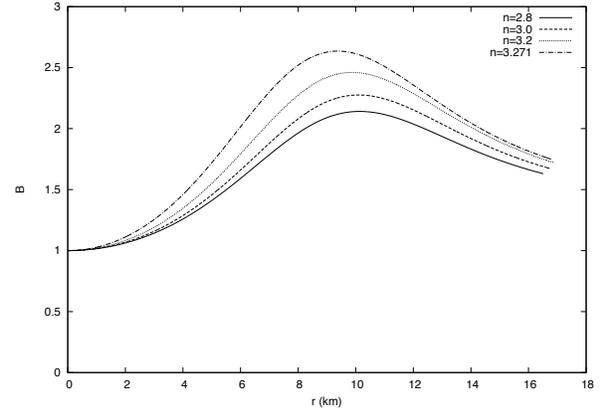}} \\
  \caption{The metric function ${\bf B}(r)$, for different solutions close to the maximum mass solution.}
  \label{Old}
 \end{center}
\end{figure} 

\begin{figure}[h]
 \begin{center} \hspace*{-1.5cm}
  \subfloat[Energy density for ${\cal U}_\pi$]{\includegraphics[width=0.53\textwidth]{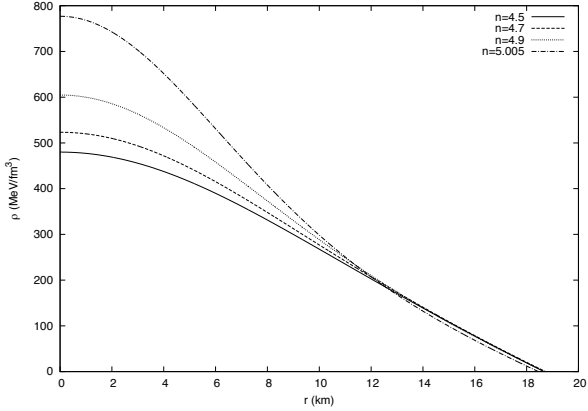}} \quad
  \subfloat[Energy density for ${\cal U}_\pi^2$]{\includegraphics[width=0.53\textwidth]{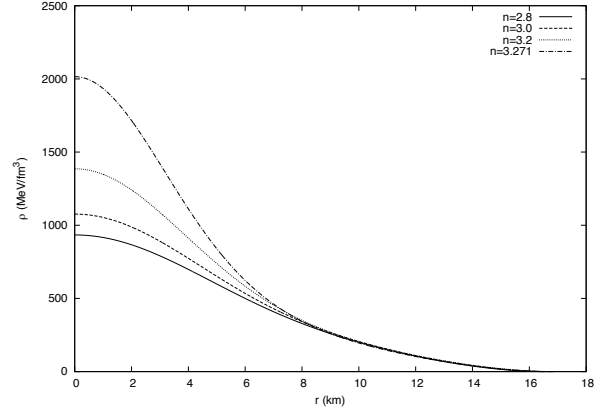}} \\
  \caption{The energy density $\rho(r)$, for different solutions close to the maximum mass solution.}
  \label{Old}
 \end{center}
\end{figure}

\section{Discussion}
We used the BPS Skyrme model (\ref{BPS-en-funct}) for the description of neutron stars and found that by simply fitting the two model parameters to the nucleon mass and radius we already get very reasonable results for the resulting neutron star masses and radii. In particular, for the maximum possible neutron star mass we find $M_{\rm max} = 2.44 M_\odot$ or $M_{\rm max} = 3.73 M_\odot$, respectively,  for the two potentials considered. This compares extremely well with the observational constraint $M_{\rm max} \sim 2.5 M_\odot$. We take this, together with the perfect fluid behaviour of the model,  as a further very strong indication that, indeed, the BPS Skyrme model provides the most important contribution to the static bulk properties of nuclear matter. In a strict sense, our results are not yet final predictions of neutron star properties, because genuine predictions require the knowledge of the full near-BPS Skyrme model (\ref{near-BPS}) together with the values of all its coupling constants, which should follow from an application to nuclear physics and the corresponding detailed fit to nuclear data. The full near-BPS Skyrme model  may also lead to a further improvement in the description of neutron stars, in the following sense. Even if the additional (standard Skyrme) terms are quite unimportant in the bulk, this is not true at the surface, because at the surface the Skyrme field is close to its vacuum value, and the term ${\cal L}_6$ approaches the vacuum much faster than the standard Skyrme model terms. The standard Skyrme model is known to prefer crystalline structures for large $B$, so crystalline structures ("neutron star crust") can be expected at the surface of a neutron star described by the near-BPS Skyrme model, whereas the bulk and core remain in a fluid phase. But precisely this structure is expected in current models of neutron stars (see, e.g., \cite{piek}). 

When compared with other, more traditional methods of nuclear physics, the advantage of the (near-) BPS Skyrme model at this moment is not so much its ability to make quantitative predictions - although this, too,  should change with more detailed investigations and with advanced numerical methods, assisted by a rigorous analytical control which follows from the integrability properties of the BPS model. After all, the methods and models of nuclear physics are well developed and lead to very precise descriptions of nuclei and nuclear matter.  However, a drawback of many models of nuclear physics is that they are tailor-made to describe rather specific physical phenomena, therefore it is difficult to use them for extrapolations to new phenomena or parameter values where they have not been employed before. We think it is one of the outstanding features of the BPS Skyrme model that it captures a {\em generic} property of (bulk) nuclear matter and allows, therefore, for far-reaching extrapolations. Concretely, in the present letter we extrapolated from $B=1$ (which provided the parameter fit values)  to $B\sim 10^{57}$ (the neutron star) and from a nonrelativistic to a highly relativistic regime, with very accurate results. In other words, the (near) BPS Skyrme model provides a {\it unified} description of nuclear matter, reaching from nucleons and atomic nuclei to neutron stars. 

There are two particular (related) results of our calculations which are somewhat different from most traditional nuclear physics calculations of neutron stars using the TOV equations (\ref{eq1}), (\ref{eq2}), although they are completely compatible with all observational data. In the traditional approach, the metric function ${\bf B}(r)$, the energy density $\rho (r)$ and the pressure $p(r)$ are considered as independent field variables, so the two TOV equations (\ref{eq1}), (\ref{eq2}) must  be closed by a third equation. For this, usually a universal algebraic equation of state (EoS) $p=p(\rho)$ resulting from the thermodynamic limit of a nuclear effective field theory (EFT) (like Quantum Hadron Dynamics (QHD) \cite{walecka}) is assumed. In our model, on the other hand, we find that already the EFT itself is of the perfect-fluid type defining its own energy density and pressure, both of which depend on the metric in an explicit fashion. It is, therefore, not possible to define a universal, algebraic off-shell EoS, and the true off-shell EoS relating $\rho$ and $p$ is a complicated and metric-dependent differential equation. 
We remark that our off-shell EoS share some features with the "quasi-local" EoS explicitly depending on the geometry (e.g., metric or curvature), which were introduced in \cite{visser} for the description of anisotropic stars and further studied in \cite{horvat} and, in relation with neutron stars, in \cite{berti}.  It turns out that in stars with anisotropic matter such quasi-local EoS are even required for consistency \cite{visser}. 
In our case, it is still possible to find (numerically) an on-shell algebraic EoS for solutions $\rho (r)$ and $p(r)$, but this on-shell EoS is no longer universal and depends on the neutron star mass or baryon number $B$. This does {\em not} mean that the EoS of nuclear matter depends on the sample size. The EoS for the BPS Skyrme model without gravity is always the same, $p=0$ at equilibrium (nuclear saturation), for arbitrary $B$. The $B$ dependence of the on-shell EoS for self-gravitating nuclear matter in the BPS Skyrme model is {\em exclusively} a consequence of self-gravitation.  Due to the nonlinearity of gravity, the effects of self-gravitation are stronger for larger $B$ (larger neutron star mass) and the effective on-shell EoS, therefore, gets stiffer. Concretely, we found an effective on-shell EoS of the type $p=a(B)\rho^{b(B)}$, see Eq. (\ref{EoS}), where $a(B)$ increases with increasing $B$ whereas $b(B)$ decreases. 

This increasing stiffness has a particular physical effect in the cases we considered, namely a neutron star radius $R$ which grows with the neutron star mass $M$, i.e., $(dM/dR)>0$ (except for stars very close to their maximum mass in the case of the potential ${\cal U}_\pi$, see Fig. 1). This behavior is at variance with the results found for solutions of the TOV equations for many (fixed, universal) nuclear physics EoS, where the neutron star radius is either essentially constant for a range of neutron star masses or even shrinks with increasing mass \cite{latt}. The reason for this behavior is that for a fixed EoS the increasing strength of self-gravitation for larger masses may collapse the star to much higher densities and, for softer EoS, even to smaller sizes. Only sufficiently stiff universal EoS are compatible with $(dM/dR)>0$. We remark that one particular case of an EoS which is sufficiently stiff to support $(dM/dR)>0$ for almost all values of $M$ is precisely given by the Skyrme crystal of Ref. \cite{piette3}. The $M(R)$ curve found there is, in fact, quite similar to the one we find for the pion mass potential ${\cal U}_\pi$, see Fig. 1b. 
In the BPS Skyrme model,  the squeezing effect of nonlinear self-gravitation is balanced by the increasing stiffness of the on-shell EoS. We emphasize that, at present, $(dM/dR)>0$ is compatible with observations and that the observational data are not yet sufficiently precise to settle this question. If $(dM/dR)>0$ finally turns out to be true, this either rules out a large class of EoS which are well motivated from nuclear physics, because only very stiff fixed EoS are compatible with $(dM/dR)>0$. Or it may indicate that in the traditional derivation of the EoS from an EFT like QHD one has to go beyond mean field theory, such that backreaction effects of gravity on the EoS may be taken into account, at least for nuclear matter in sufficiently strong gravitational fields. A detailed discussion of these issues will be given elsewhere. In any case, the qualitative results we found for the EoS within the BPS Skyrme model also point towards possible improvements of the standard nuclear physics approach to neutron stars in strong gravitational fields. 

There are many ways in which the present investigation can be deepened and extended. One obvious possibility is to use additional potentials and to study how the shapes of these potentials influence the properties of the resulting neutron stars, e.g., which maximal masses can be reached and for which potentials the relation $(dM/dR)>0$ remains true. Another interesting research direction is related to rotating neutron stars and to neutron stars in magnetic fields. In principle, both these tasks are rendered feasible by the fact that it is known how to rotate Skyrmions (for a recent discussion see, e.g., \cite{krusch}) and what is the correct, QCD induced coupling of Skyrmions to the electromagnetic interaction \cite{witten2}. Still, the resulting systems are no longer spherically symmetric, so a full system of PDEs has to be solved numerically in these cases. A further step in the analysis would be to use the full near-BPS Skyrme model as a basis for the calculation of neutron star solutions and properties, but here, in a first step, the detailed application of the near-BPS Skyrme model without gravity to nuclei and nuclear matter is required. As the BPS and integrability properties are no longer available in this case, full three-dimensional numerical calculations will be necessary.

\section*{Acknowledgement}
The authors acknowledge financial support from the Ministry of Education, Culture and Sports, Spain (grant FPA2011-22776), 
the Xunta de Galicia (grant INCITE09.296.035PR and
Conselleria de Educacion), the
Spanish Consolider-Ingenio 2010 Programme CPAN (CSD2007-00042), and FEDER. 
CN thanks the Spanish Ministery of
Education, Culture and Sports for financial support (grant FPU AP2010-5772).
Further, AW was supported by polish NCN (National Science Center) grant DEC-2011/01/B/ST2/00464 (2012-2014). 
JSG thanks M. A. Perez-Garcia for discussions.


\begin{thebibliography}{99}

\bibitem{skyrme} T. H. R. Skyrme, Proc. Roy. Soc. Lon. {\bf 260},
127 (1961); Nucl. Phys. {\bf 31}, 556 (1962); J. Math. Phys. {\bf
12}, 1735 (1971).
\bibitem{thooft} G. t'Hooft, Nucl. Phys. B{\bf 72}. 461 (1974); E. Witten, Nucl. Phys. B{\bf 160}, 57 (1979); E. Witten, Nucl. Phys. B{\bf 223}, 433 (1983).
\bibitem{wood}
R. A. Battye, N. S. Manton, P. M. Sutcliffe, S. W. Wood,
Phys. Rev. C{\bf 80} (2009) 034323. 
\bibitem{lau}
P. H. C. Lau, N. S. Manton,
e-Print: arXiv:1408.6680 [nucl-th].
\bibitem{manton-book}
N. Manton, P. Sutcliffe, "Topological Solitons", Cambridge University Press, Cambridge, 2007.
\bibitem{crystal1}
M. Kugler, S. Shtrikman,
Phys. Lett. B{\bf 208} (1988) 491.
\bibitem{crystal2}
L. Castillejo, P. S. J. Jones, A. D. Jackson, J. J. M. Verbaarschot, A. Jackson,
Nucl. Phys. A{\bf 501} (1989) 801.  
\bibitem{BPS-Sk}
C. Adam, J. Sanchez-Guillen, A. Wereszczynski,
Phys. Lett. B{\bf 691}, 105 (2010);
C. Adam, J. Sanchez-Guillen, A. Wereszczynski,
Phys. Rev. D{\bf 82}, 085015 (2010).  
\bibitem{sutcliffe}
P. Sutcliffe, JHEP {\bf 1008}, 019 (2010); JHEP {\bf 1104}, 045 (2011).
\bibitem{bind}
C. Adam, C. Naya, J. Sanchez-Guillen, A. Wereszczynski,
Phys. Rev. Lett. {\bf 111} (2013), 232501;
C. Adam, C. Naya, J. Sanchez-Guillen, A. Wereszczynski,
Phys. Rev. C{\bf 88} (2013), 054313.  
\bibitem{marleau}
E. Bonenfant, L. Marleau,
Phys. Rev. D{\bf 82}, 054023 (2010);
E. Bonenfant, L. Harbour, L. Marleau,
Phys. Rev. D{\bf 85}, 114045 (2012); M.-O. Beaudoin, L. Marleau, Nucl. Phys. B{\bf 883} (2014) 328.
\bibitem{bizon}
P. Bizon, T. Chmaj,
Phys. Lett. B{\bf 297} (1992) 55.
\bibitem{piette1}
B. M. A. G. Piette, G. I. Probert,
Phys. Rev. D{\bf 75} (2007) 125023.
\bibitem{piette2}
S.G. Nelmes, B. M. A. G. Piette,
Phys. Rev. D{\bf 84} (2011) 085017. 
\bibitem{walhout}
T. S. Walhout,
Nucl. Phys. A{\bf 484} (1988) 397;
T. S. Walhout,
Nucl. Phys. A{\bf 519} (1990) 816.
\bibitem{piette3} 
S. G. Nelmes, B. M. A. G. Piette,
Phys. Rev. D{\bf 85} (2012) 123004. 
\bibitem{nelmes-thesis}
S. G. Nelmes, "Skyrmion Stars", Durham Theses, Durham University 2012, available online at: http://etheses.dur.ac.uk/5258/.
\bibitem{fosco}
C. Adam, C. D. Fosco, J. M. Queiruga, J. Sanchez-Guillen, A. Wereszczynski,
J. Phys. A{\bf 46}, 135401 (2013). 
\bibitem{speight}
J. M. Speight, 
J. Phys. A{\bf 43}, 405201 (2010). 
\bibitem{thermo}
C. Adam, C. Naya, J. M. Speight, J. Sanchez-Guillen, A. Wereszczynski, 
Phys. Rev. D{\bf 90} (2014) 045003.
\bibitem{slobo}
R. Slobodeanu, Int. J. Geom. Methods Mod. Phys. {\bf 08}, 1763 (2011).
\bibitem{speight2}
J. M. Speight, arXiv:1406.0739.
\bibitem{weinberg}
S. Weinberg, "Gravitation and Cosmology", Wiley, New York, 1972.
\bibitem{latt}
J. M. Lattimer,
Ann. Rev. Nucl. Part. Sci. {\bf 62} (2012) 485; 
A. W. Steiner, J. M. Lattimer, E. F. Brown,
Astrophys. J. {\bf 765} (2013) L5;
K. Hebeler, J. M. Lattimer, C. J. Pethick, A. Schwenk,
Astrophys. J. {\bf 773} (2013) 11;
J. M. Lattimer, A. W. Steiner,
Astrophys. J. {\bf 784} (2014) 123. 
\bibitem{ozel}
F. \"Ozel, G. Baym, T. G\"uver, Phys. Rev. D{\bf 82}, 101301 (2010);
F. Ozel, D. Psaltis, R. Narayan, A. Santos Villarreal,
Astrophys. J. {\bf 757} (2012) 55;
T. Guver, F. Ozel,
Astrophys. J. {\bf 765} (2013) L1.  
\bibitem{piek}
H. Horowitz, J. Piekarewicz,
Phys. Rev. Lett. {\bf 86}, 5647 (2001); J. Piekarewicz, J. Phys. Conf. Ser. {\bf 492} (2014) 012008; 
C. J. Horowitz, M. A. Perez-Garcia, D. K. Berry, J. Piekarewicz,
Phys. Rev. C{\bf 72} (2005) 035801.
\bibitem{walecka}
J. D. Walecka, Ann. Phys. {\bf 83} (1974) 491; B. D. Serot, J. D. Walecka, Int. J. Mod. Phys. E{\bf 6} (1997) 515.
\bibitem{visser}	
C. Cattoen, T. Faber, M. Visser,
Class. Quant. Grav. {\bf 22} (2005) 4189; M. Visser, PoS BHGRS (2008) 001.
\bibitem{horvat}
D. Horvat, S. Ilijic, A. Marunovic,
Class. Quant. Grav. {\bf 28} (2011) 025009.
\bibitem{berti}
H. Silva, C. Macedo, E. Berti, L. Crispino,
arXiv:1411.6286. 
\bibitem{krusch}
R. A. Battye, M. Haberichter, S. Krusch,
e-Print: arXiv:1407.3264 [hep-th].
\bibitem{witten2}
C. G. Callan, E. Witten, 
Nucl. Phys. B{\bf 239} (1984) 161. 

\end{thebibliography}
\end{document}